\documentclass{aa}  

\usepackage{graphicx}
\usepackage{subcaption}
\usepackage[colorlinks=true, citecolor=blue, linkcolor=blue, urlcolor=black]{hyperref}
\usepackage{txfonts}

\begin{document}

   \title{Re-evaluation of Lunar X-ray observations by Apollo 15 \& 16\thanks{This work is dedicated to Al(fred Merrill) Worden $\dagger$, who operated the XRFS instrument during the Lunar orbit of the Apollo 15 mission.}}

   \author{A. J. Gloudemans \inst{1,2,3}
   \and E. Kuulkers\inst{1}
   \and R. Campana\inst{4}
   \and A. Escalante\inst{5}
   \and M. Kole\inst{6}
   \and Y. Mollard\inst{7}
   } 
    \institute{ESA/ESTEC, Keplerlaan 1, 2201 AZ Noordwijk, The Netherlands \label{inst1} 
    \and Anton Pannekoek Institute for Astronomy, University of Amsterdam, Science Park 904, 1098 XH, Amsterdam, The Netherlands\label{inst2}
    \and Leiden Observatory, Leiden University, PO Box 9513, 2300 RA Leiden, The Netherlands \\ e-mail: gloudemans@strw.leidenuniv.nl \label{inst7}
    \and INAF/OAS, Via Gobetti 101, I-40129, Bologna, Italy\label{inst3}
    \and ESA/ESAC, Camino Bajo del Castillo, 28692 Villafranca del Castillo, Madrid, Spain\label{inst4}
    \and Department of Nuclear and Particle Physics, University of Geneva, 24 quai Ernest-Ansermet, 1205, Geneva, Switzerland\label{inst5}
    \and Bordeaux INP, Avenue des facult\'es, 33405 Talence, France\label{inst6}
    \\}
   \date{Received 12 January 2021 / Accepted 16 April 2021}

  \abstract{The Apollo 15 \& 16 missions were the first to explore the Lunar surface chemistry by investigating about 10\% of the Lunar surface using a remote sensing X-ray fluorescence spectrometer experiment. The data obtained have been extensively used to study Lunar formation history and geological evolution. In this work, a re-evaluation of the Apollo 15 \& 16 X-ray fluorescence experiment is conducted with the aim of obtaining up-to-date empirical values for aluminum (Al) and magnesium (Mg) concentrations relative to silicon (Si) of the upper Lunar surface. An updated instrument response, a newly reconstructed Lunar trajectory orbit, and improved intensity ratio calculations were used to obtain new intensity ratio maps. The resulting Lunar Al/Si and Mg/Al X-ray maps show a clear distinction in Lunar mare and highland regions. The mean Al/Si and Mg/Al intensity ratios for the mare regions obtained from the newly obtained maps are $0.54\pm0.07$ and $0.54\pm0.17$, respectively; for the highland regions, the values are $0.76\pm0.07$ and $1.07\pm0.13$, respectively. For the Mg/Si intensity ratio, no clear distinction between Lunar features is obtained and we derived a mean value of $0.47\pm0.13$. Our determined intensity ratios are lower than previously published. These values can be used to infer concentration ratios when accounting for Solar activity, inter-orbit variability, and measurements from different instruments. We employed a correction to infer concentration ratios by comparing our intensity ratios directly to Lunar rock concentrations obtained from various Lunar missions.}


   \maketitle


\section{Introduction}

The Apollo 15 \& 16 spacecrafts were launched on July 26, 1971 and April 16, 1972, respectively. The Lunar orbiting spacecraft (Command Service Module, CSM) on both missions carried out an X-ray fluorescence spectrometer (XRFS) experiment (\citealt{adler1972apollo, adler1975apollo, jagoda1974apollo}) within the Scientific Instrumentation Module (SIM), which was used for Lunar surface exploration. X-ray spectroscopy is of high importance for geochemical exploration of the Moon and other bodies, such as planets and asteroids, as it can be used to study extended surface areas, without having to land on their surfaces. X-rays can be observed via fluorescence and scattering of Solar X-rays on a body's surface. Solar X-rays reaching the Lunar surface excite electrons in medium to high atomic number elements (such as aluminum), which in turn emit characteristic X-rays from the specific element. The chemical composition of the surface can give information on, for example, the formation history of the body or crater impact events. The Apollo 15 \& 16 missions were the first to use an XRFS experiment in orbit (see, e.g., \citealt{clark1997remote}) and also the first to investigate a large part ($\sim$10\%) of the Lunar surface. These measurements were the basic ingredient of our early understanding of the geology and formation of the Lunar surface.
Also more recent Lunar missions such as SMART-1/D-C1XS \citep{Racca2001EM&P...85..379R, FOING20066}, Chandryaan 1/C1XS \citep{narendranath2011lunar}, and Chang'E-2/CE2XRS \citep{Dong2016RAA....16....4D} have since used orbital X-ray fluorescence spectroscopy to study the Lunar surface. Moreover, X-ray telescopes, such as ROSAT \citep{schmitt1991Natur.349..583S}, ASCA \citep{Kamata1999AdSpR..23.1829K}, and Chandra \citep{wargelin2004ApJ...607..596W}, have also observed the Lunar surface. 
In addition to this, fluorescence spectroscopy has been used in studies of, for example, Mercury by MESSENGER/XRS \citep{schlemm2007SSRv..131..393S, Nittler2011Sci...333.1847N, Weider2015E&PSL.416..109W} and other Solar system moons and asteroids (e.g., 433 Eros, \citealt{nittler2001x}, and asteroid Itokawa, \citealt{Okada2007LPI....38.1287O}).

In this work, a re-evaluation of the Apollo 15 \& 16 XRFS experiment data is performed using an updated instrument response, a newly reconstructed Lunar trajectory orbit, and improved intensity ratio calculations. The data are not only of historical value, but doing the re-evaluation now also enables us to compare the results with those of much later X-ray fluorescent experiments (such as with the Chandryaan-1/C1XS) and up-to-date Lunar sample analyses. 
Our primary aim is to create chemical intensity ratio maps of measured elements, such as aluminum (Al) and magnesium (Mg) with respect to silicon (Si), on the Lunar surface using newly obtained values and make a comparison with inferred elemental concentrations found in Lunar rocks.

\section{Apollo 15 and 16 X-ray observations}

\begin{table*}
\caption{Log of the observations with the XRFS onboard Apollo 15 \& 16 in normal and extended mode of Detector \#1.}      
\label{tab:observations}      
\centering  
\resizebox{\textwidth}{!}{
\begin{tabular}{c c | c | c | c }  
\hline\hline       
 & & Date start observation (UTC) & Date end observation (UTC) & Total exposure time (h)\\
\hline
\textbf{Apollo 15} & Normal mode & 30 Jul 1971 02:44:07 &  04 Aug 1971 18:21:48 & 54.62 \\
          & Extended mode &  30 Jul 1971 08:09:10 &  04 Aug 1971 12:40:55 & 14.78  \\
\hline
\textbf{Apollo 16} & Normal mode &  20 Apr 1972 01:58:03 &  24 Apr 1972 17:47:31 & 50.00  \\
          & Extended mode &  20 Apr 1972 07:45:15 &  24 Apr 1972 19:24:22 & 14.52 \\
\hline \hline
\end{tabular}}
\end{table*}

\begin{table*}
\caption{Absorption cross section, density, and filter thickness of the Mg, Al, and Be filter at K$\alpha$ line energies of Mg, Al, and Si. The filter thickness is adopted from \cite{yin1993x}. Absorption cross sections were determined using the web tool \textit{xraylib} \citep{SCHOONJANS2011776}. }
\label{tab:var_trans}      
\centering  
\begin{tabular}{c c c l l l }  
\hline\hline       
 & \multicolumn{3}{c}{Absorption cross section \textbf{$\mu$} ($cm^2 g^{-1}$)} & \textbf{$\rho$} ($g$ $cm^{-3}$) & $d$ ($\mu$m)\\ 
\hline
Element & 1.254 keV (Mg) & 1.487 keV (Al) & 1.740 keV (Si) & & \\ \hline
Be & 307.76 & 184.47 & 114.47 & 1.84 & 25.4 \\ 
Mg & 502.39 & 4084.09 & 2782.42 & 1.74 &  5.1 \\ 
Al & 648.13 & 411.68 & 3111.63 & 2.70 &  6.4 \\ \hline
\hline
\end{tabular}
\end{table*}

\subsection{XRFS instrument}

The XRFS instrument consisted of three detectors and a Solar monitor, which were all proportional counters. The proportional counters were filled with 90$\%$ argon (Ar), 9.5$\%$ carbon dioxide (CO$_2$), and 0.5$\%$ helium (He). All three detectors had a beryllium (Be) window with a thickness of approximately 25 $\mu$m (e.g., \citealt{adler1972apollo}) and each detector had an effective window area of about 25 cm$^2$.

One detector had no filter (referred to as Detector \#1), whereas the other two detectors had an Mg (Detector \#2) and an Al filter (Detector \#3), respectively, in front of their windows. The thickness of the Mg and Al filters, as well as that of the Be windows, is not entirely clear from the literature as different values have been published (compare, e.g., \citealt{adler1972apollo, adler1975apollo, yin1993x}). An overview of these different values and the implication thereof is discussed in Appendix \ref{sec:appendix_filterthickness}. 
We assume that the Mg and Al foils have a thickness of 5.1 and 6.4 $\mu$m, respectively (e.g., \citealt{yin1993x}).
The different filters provided selective filtering of the elements Si, Al, and Mg (see, e.g. \citealt{clark1997remote}) since the energy resolution of the individual proportional counters could not adequately separate the Si, Al, and Mg lines. The field-of-view (FOV) of the detectors were set by collimator cells placed in front of the three detectors, which resulted in a FOV of approximately $\pm 30\degr$  at full-width at half maximum (see \citealt{adler1975apollo}). A sample analysis shows the collimator is presumably made of Cu-Zn alloy (Kuulkers et al. in prep).

The data handling system recorded the X-ray photons in eight spectral channels. The XRFS was able to operate in a normal mode and extended mode. In normal mode, the first seven channels of all detectors covered an energy range of 0.69--3.0 keV. Channel 8 covered all energies higher than 3.0 keV \citep{jagoda1974apollo}. In the extended mode, each channel energy range of Detector 1 was doubled. In both modes, each detector output was read out every 8 seconds. The gain, resolution, and efficiency of the detectors were calibrated by using in-flight calibration sources consisting of radioactive sources $^{55}$Fe and Mn-K$_\alpha$. Calibration measurements were done every 16 minutes for a duration of 64 seconds. 

The Solar monitor simultaneously measured the Sun's X-ray output with the detectors observing the Lunar surface: It was placed at the other side of the CSM with respect to the three detectors. The Apollo 15 Solar monitor had one Be window, while the Apollo 16 Solar monitor had two Be windows. The extra Be filter was added for the Apollo 16 mission because the detector experienced gain shifts during the Apollo 15 flight \citep{apollo16preliminary}. For further details on the XRFS instrument, we refer readers to \cite{adler1972apollo}, \cite{jagoda1974apollo}, and \cite{adler1975apollo}. 

\subsection{Lunar surface observations}

The Apollo 15 and 16 XRFS observations were performed during 5.7 and 4.7 days in Lunar orbit, respectively (see Tab.\ \ref{tab:observations}), and together they measured $\sim$10\% of the Lunar surface (see, e.g., \citealt{yin1993x}). The XRFS instrument operated in normal mode for $\sim$75 \% of the time and the remaining time in extended mode. 
We retrieved the Apollo 15 and 16 X-ray data from the Planetary Data System (PDS) at the NASA Space Science Data Coordinated (NSSDC) Archive.
For each mission, the data sets contain ASCII tables of time-ordered, raw event count rates acquired by the XRFS from July 30 to August 4, 1971 (Apollo 15; \citealt{adler2017}) and from April 20 to 24, 1972 (Apollo 16; \citealt{trombka2017}) during Lunar orbit.

The Apollo 15 trans-Earth coast XRFS data as well as the full Apollo 15 and 16 merged data sets from the Gamma-Ray Spectrometer (GRS; \citealt{harrington1974}) were available as restored binary offline tape data sets at the NSSDC Archive and deciphered by us. The trajectory information stored in the data from the GRS was used to reconstruct, with high accuracy, the orbit of the CSM, propagating the positions and velocities expressed in the inertial frame B1950, centered on the Moon. 
Given the CSM position and velocity computed from the propagation of the GRS data, the transformation from the B1950 inertial reference frame to the local horizontal reference frame can be computed at the times of the XRFS measurements for which the body angles are also known. Then, using the Euler angles from the XRFS tape to transform from local horizontal to body-fixed, the attitude of the CSM is unequivocally defined.

During the Lunar orbit, the CSM continuously changed altitude. The CSM was in a low-altitude orbit ($\sim$20--100 km) during the first part of both Apollo missions and in a higher-altitude orbit ($\sim$100--120 km) during the second part.

\subsection{Data processing} 
\label{sec:dataprocessing}

Intensity ratios of measured Al, Si, and Mg intensities were used to eliminate effects of non-geochemical variations such as the Solar illumination angle and Lunar surface roughness (e.g., \citealt{adler1972apollo}). These intensity ratios are Al/Si, Mg/Si, and Mg/Al. The Si abundance on the Lunar surface is high and stable, changing only by a maximum of 5\%\ across the surface, hence the ratios are mostly a measure of the Al and Mg variation, and concentration, on the Lunar surface (e.g., \citealt{yin1993x}). 

We made multiple assumptions when examining the data. Firstly, 100\% efficiency of the proportional counter was assumed, that is, all X-rays that enter the proportional counter are absorbed by the argon gas and give rise to an X-ray detection (see e.g., \citealt{yin1993x}). Secondly, it was assumed that the three detectors have identical characteristics except for their filters (see, e.g., \citealt{adler1972apollo}). For example, the seven spectral channels of each detector of both Apollo missions cover the same energy range of 0.69--3.0 keV in normal mode \citep{jagoda1974apollo}.

In the first step of data processing, the data obtained during calibration periods were removed (see also \citealt{adler2017, trombka2017}). The contribution due to background radiation was determined from the XRFS observations in Lunar orbit 
$\sim$30--150 degrees away from the sub-Solar point. The background was determined for each detector and each spectral channel independently and subtracted from the detector channel count rates.

During Apollo 16, the Solar spectrum was softer than during the Apollo 15 mission \citep{bielefeld1977lunar}. The softening of the Solar spectrum caused a decrease in excitation of Si compared to Al and Mg during Apollo 16 since Si has a higher atomic number. We corrected the Apollo 16 observations for that effect (see Sect.~\ref{chemistry}). Also, during several orbits of both missions, the Solar activity was higher than usual (e.g., revolution 67 of Apollo 15, see \citealt{adler1972apollo}), causing enhanced excitation of Si. These orbits were removed from the data in our analysis (see Sect.~\ref{chemistry}).

To determine the characteristic line intensities of Mg, Al, and Si, the transmission as a function of energy of each filter has to be taken into account. The expression for transmission can be written as follows (see \citealt{adler1972apollo}):
\begin{equation}
\label{eq:transmission}
    G_{i} = \sum_{j=1}^3 I_j^T T_j^{Be}F_{ij},
\end{equation}
where $G_i$ is the total photon counts in detector $i$; $I_j^T$ is the total X-ray intensity at energy $j$ for the Mg, Al, and Si line; $T_j^{Be}$ is the X-ray transmission efficiency of the Be detector window; and $F_{ij}$ are the filter transmission efficiencies of the Mg and Al filters for energy $j$.  
The intensity received by Detectors \#1, \#2, and \#3 is given by the following:
\begin{align}
    \label{eq:intensity}
    I_1 &= I'_1 e^{-(\mu_{Be} \rho_{Be} d_{Be})}, \\
    I_2 &= I'_2 e^{-(\mu_{Be} \rho_{Be} d_{Be}+\mu_{Mg} \rho_{Mg} d_{Mg})},\\
    I_3 &= I'_3 e^{-(\mu_{Be} \rho_{Be} d_{Be}+\mu_{Al} \rho_{Al} d_{Al})},
\end{align}
with $I'_i$ being the incident intensity, $\mu$ being the absorption cross section of the specific element in cm$^{-2}$g, $\rho$ being the density in g cm$^{-3}$, and $d$ being the thickness of the filter in centimeters. The transmission factors are subsequently given by $I/I'$ and can be determined at the Mg, Al, and Si K$\alpha$ lines at 1.254, 1.487, and 1.740 keV, respectively. Nowadays, the absorption cross section for different elements are accurately determined (e.g., \citealt{SCHOONJANS2011776}). We used these absorption cross-section values, which are given in Tab.\ \ref{tab:var_trans}, together with the filter thicknesses and densities in the calculations. This is in contrast to previous work (e.g., \citealt{clark1997remote}) where power laws were used to determine the mass absorption. Subsequently, a "filter matrix" M was constructed to finally obtain the intensity ratios of Al/Si, Mg/Si, and Mg/Al, which is further explained in Appendix \ref{sec:appendix_filtercalculation}. 

The collimator transmission has been described by \cite{golub1972test}, \cite{adler1975apollo}, and \cite{bielefeld1977imaging}. When re-engineering the dimensions of the various parts of the XRFS from these reports, it became apparent that the transmissions shown in these reports could not be reproduced. Also, the re-engineered values did not match what was expected from in-flight measurements of the various X-ray sources observed during the trans-Earth coast of Apollo 15. A prototype unit of the XRFS still exists; using the dimensions of the detector assembly from this unit, the results from in-flight measurements can be reproduced, and an updated collimator transmission can be derived (Kuulkers et al., in preparation). 

\begin{figure}
        \includegraphics[width=\columnwidth]{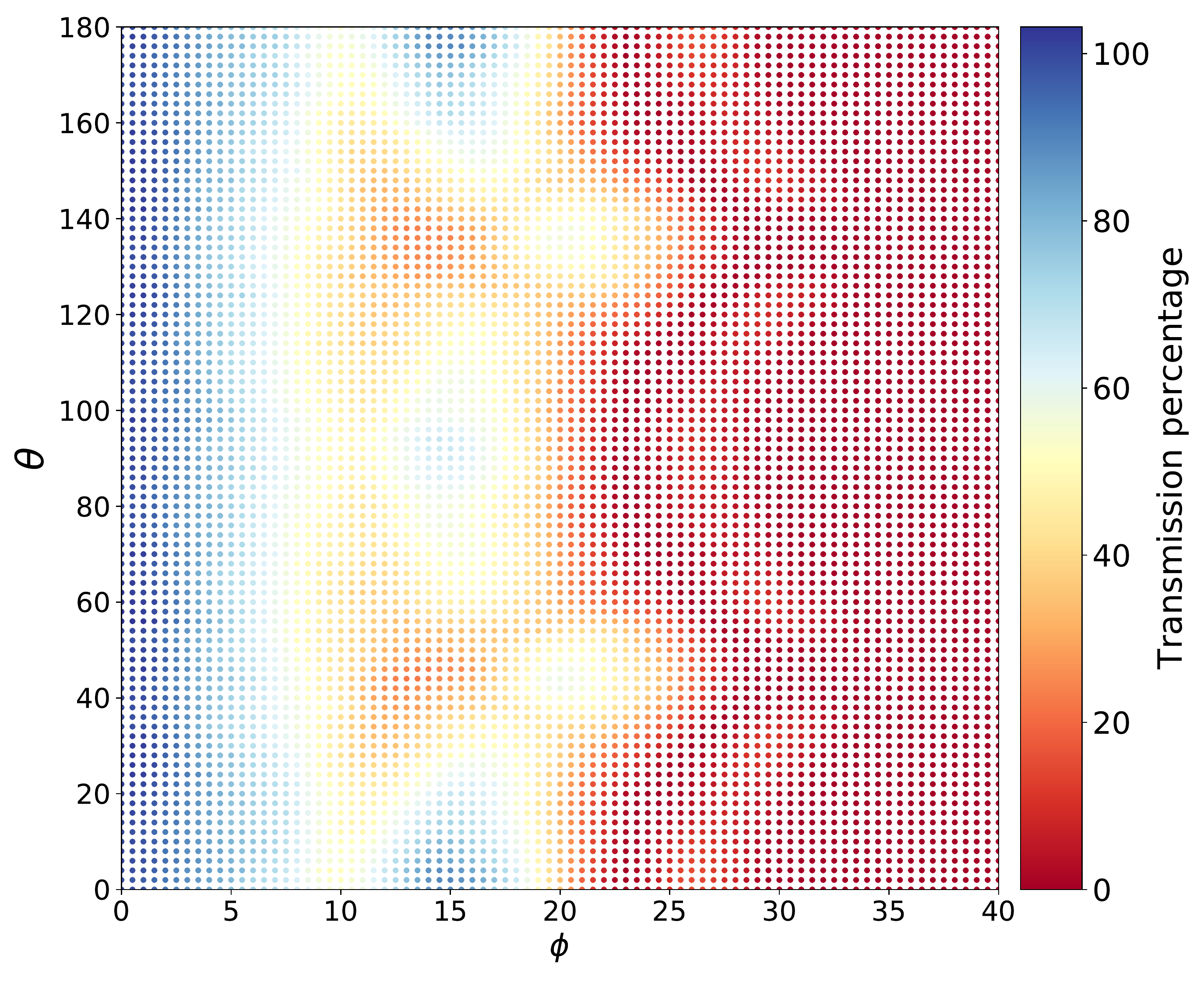}
    \caption{Collimator response (0--100\%) for angles of $0\degr<\phi<40\degr$  and $0\degr<\theta<180\degr$. The angles $\phi$ and $\theta$ represent the geometric off-axis sighting angles on the detector (see Golub et al.\ 1972, Adler et al.\ 1975).} 
    \label{fig:transmission_csm}
\end{figure}

\begin{figure*}
    \centering
    \begin{subfigure}{\textwidth}
        \centering
        \includegraphics[width=\textwidth, trim={0.0cm 0cm 0cm 0.0cm}, clip]{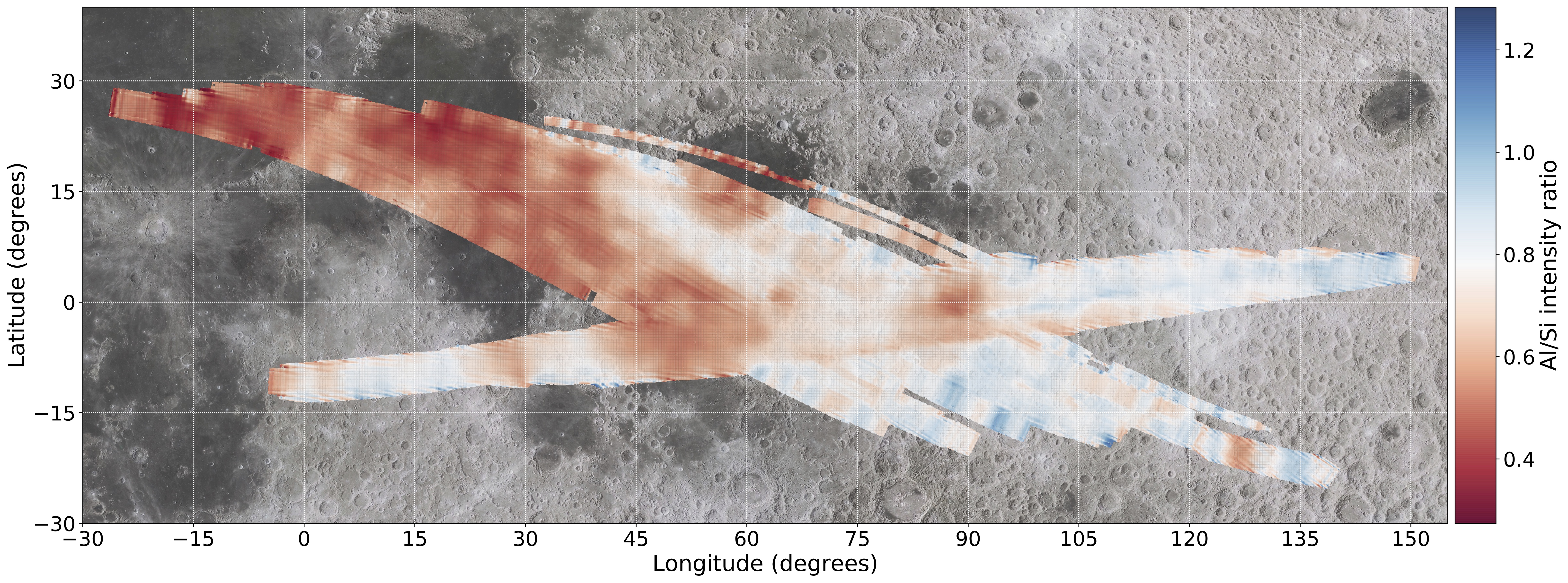}
    \end{subfigure}
     ~ 
    \begin{subfigure}{\textwidth}
        \centering
        \includegraphics[width=\textwidth, trim={0.0cm 0cm 0cm 0.0cm}, clip]{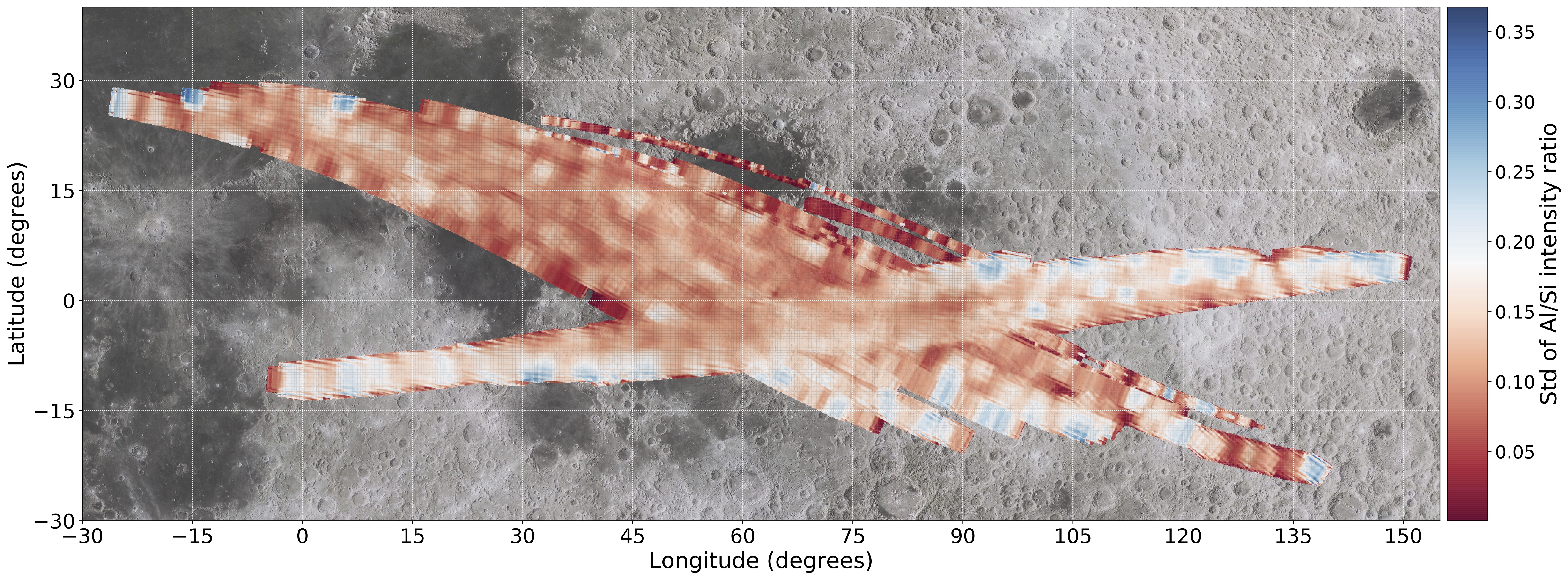}
    \end{subfigure}
    \caption{\label{fig:alsi_map} Map of Al/Si intensity ratio (top panel) and standard deviation (bottom panel) measured by the Apollo 15 and 16 XRFS projected onto an optical image of the Lunar surface. Background image credit: NASA (\url{www.Solarsystemscope.com/textures/}).}
\end{figure*}

\begin{figure}[h!]
        \includegraphics[width=\columnwidth]{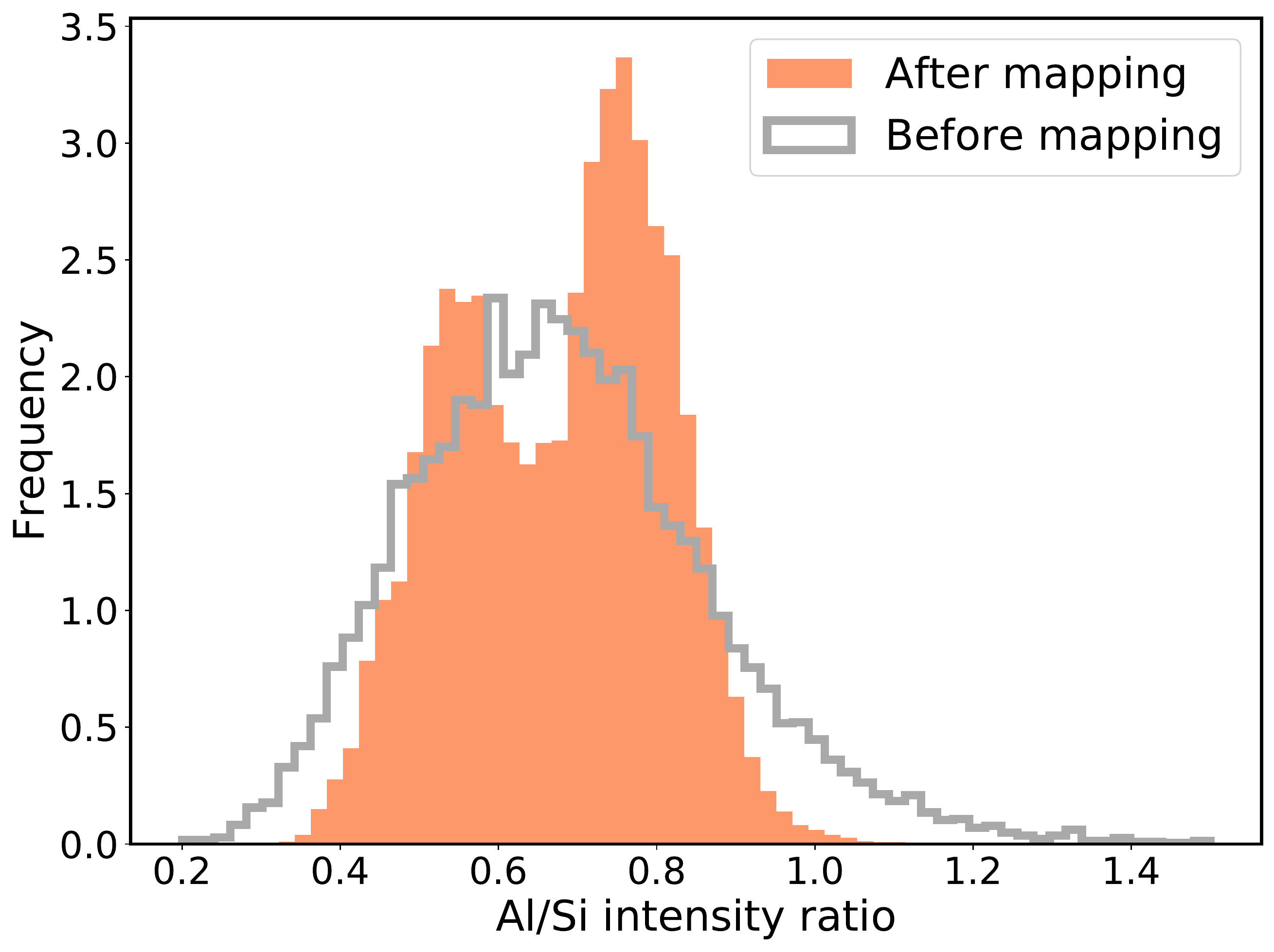}
    \caption{Corrected normalized frequency distribution of Al/Si intensity ratio before and after mapping. The mapping before shows all 8 sec Al/Si intensity ratios measured by Apollo 15 \& 16. The mapping after shows the frequency of the mean Al/Si intensity ratio stored on the constructed Lunar surface grid by taking the response into account.}
    \label{fig:double_peak_al_si}
\end{figure}

The collimator transmission values (Fig.~\ref{fig:transmission_csm}) were obtained by implementing a Monte Carlo simulation. The angles $\phi$ and $\theta$ represent the geometric off-axis sighting angles on the detector (see \citealt{golub1972test, adler1975apollo}). The Monte Carlo simulation involved a geometrical mass model of the XRFS experiment, using the Geant4 framework. Geant4 \citep{agostinelli03} is a tool kit for the simulation of the passage of particles through matter, with a large set of applications in high-energy, nuclear, accelerator, medical, and space-science physics, and it is routinely used to estimate the response of space astrophysics instrumentation. For the work presented here, version 4.10.2 of Geant4 was used with the LivermorePhysics model handling the electromagnetic process including fluorescence effects. The various parameters of the collimator and detector unit (dimensions, thicknesses) were adapted using the measurements on the prototype unit.

The response was derived by simulating $1\times10^6$ incident photons for 100 different incoming energies, ranging from 0.1 to 10 keV in 0.1 keV steps, coming from different sighting directions in the $(\theta,\phi)$ space. We used the updated collimator transmission in this work.

\begin{figure*}
    \centering
    \begin{subfigure}{\textwidth}
        \centering
        \includegraphics[width=\textwidth, trim={0.0cm 0cm 0cm 0.0cm}, clip]{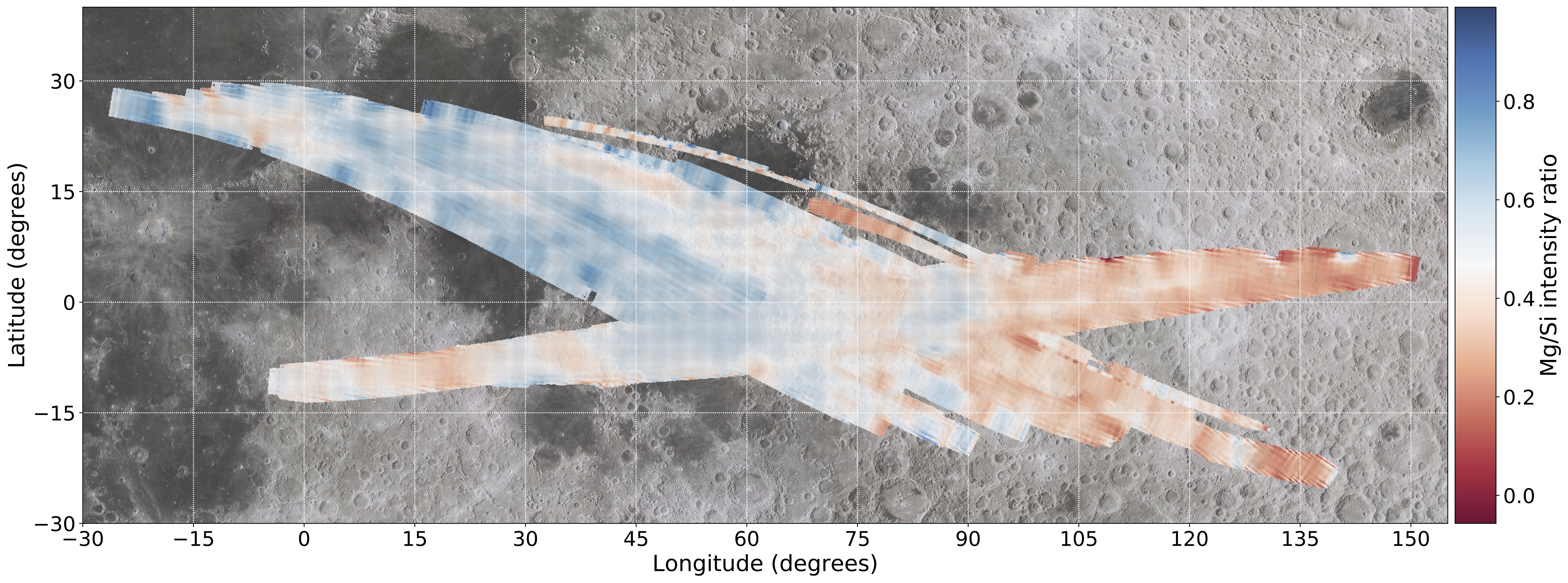}
    \end{subfigure}
     ~ 
    \begin{subfigure}{\textwidth}
        \centering
        \includegraphics[width=\textwidth, trim={0.0cm 0cm 0cm 0.0cm}, clip]{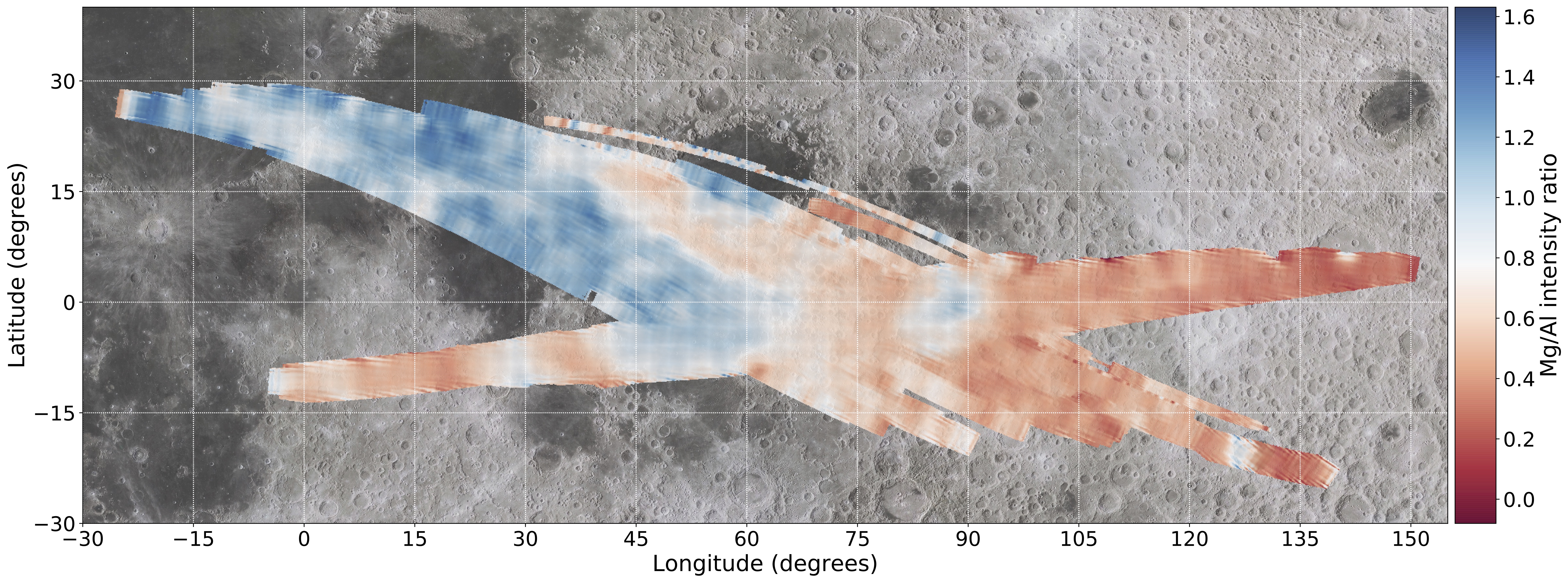}
    \end{subfigure}
    \caption{\label{fig:mg_si_mg_al_maps} Lunar surface maps of Mg/Si (top panel) and Mg/Al (bottom panel) intensity ratio measured from the Apollo 15 and 16 CSM projected onto an optical image of the Lunar surface.}
\end{figure*}

\begin{figure*}
    \centering
    \begin{subfigure}{\columnwidth}
        \centering
        \textbf{\large{Without correction}}\par\medskip
        \includegraphics[width=\textwidth, trim={0cm 0cm 0cm 0.0cm}, clip]{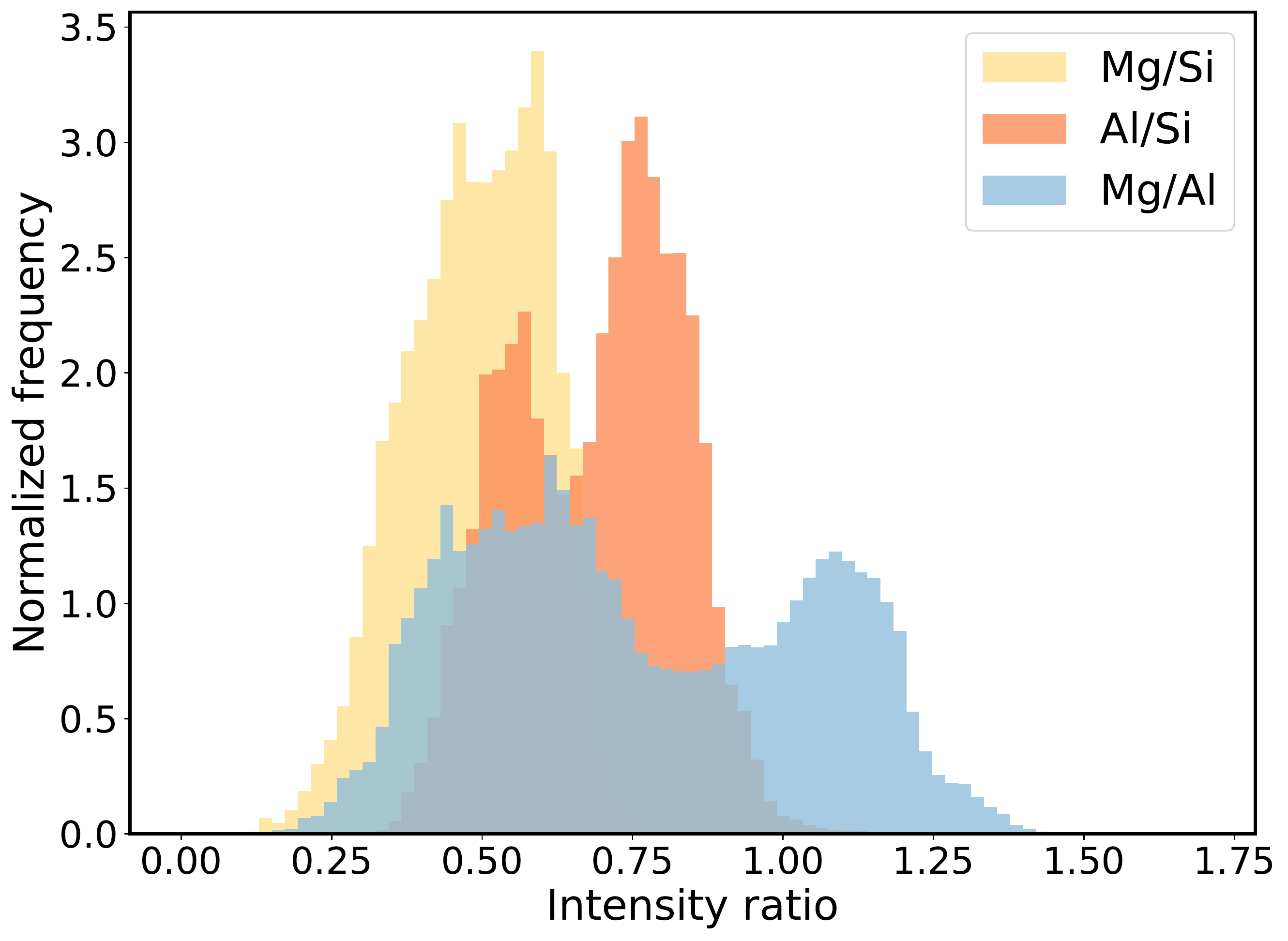}
    \end{subfigure}
     ~ 
    \begin{subfigure}{\columnwidth}
        \centering
        \textbf{\large{With correction}}\par\medskip
        \includegraphics[width=\textwidth, trim={0cm 0cm 0cm 0.0cm}, clip]{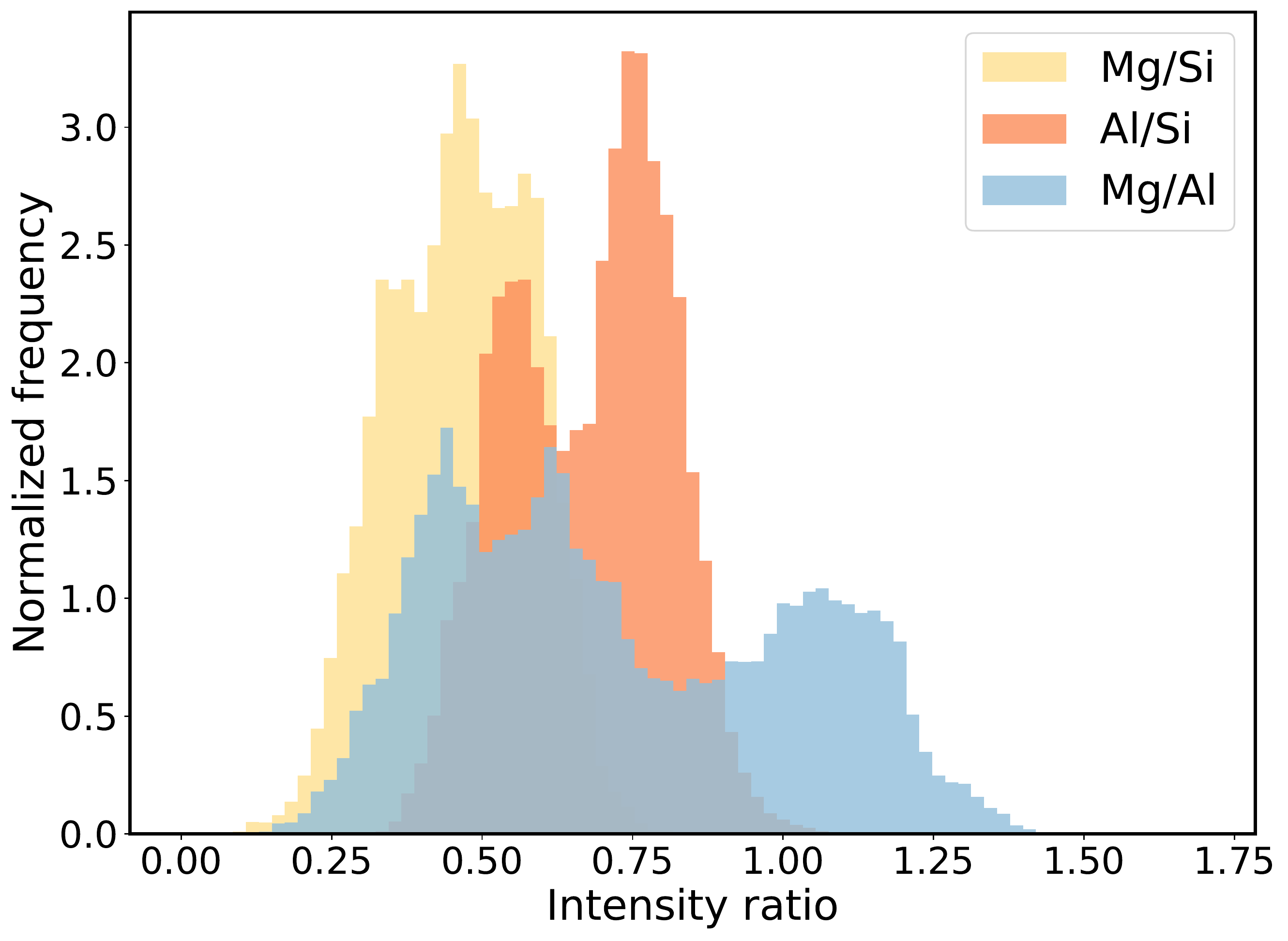}
    \end{subfigure}
    \caption{\label{fig:freq_distributions} Frequency distributions (normalized number of intensity ratios observed) of Al/Si, Mg/Si, and Mg/Al after mapping. Left: Distributions without correction of Apollo 16 data. Right: Distributions with Apollo 16 data corrected for softening of Solar spectrum. The corrections result in additional peaks appearing in the Mg/Al and Mg/Si distributions, which is due to the relatively low values measured for Mg/Si and Mg/Al of the highland area between $\sim$100--150 degrees longitude by Apollo 16. In both figures, the intensity ratio bin width is approximately 0.02.}
\end{figure*}

\subsection{Lunar surface chemistry}
\label{chemistry}

From the determined intensity ratios, geochemical maps of the surface can be reconstructed for the part of the Moon covered during each orbit. The resolution of the observations are approximately 60$\times$75 km$^2$ for the low-altitude orbits ($\sim$60 km) and 110$\times$125 km$^2$ for the high-altitude orbits ($\sim$110 km) when using the 8 sec measurements.

When constructing these maps, cuts were made to the data to remove noise and unreliable measurements. First, the intensity ratio data corresponding to the measured intensity by each detector that falls within 3$\sigma$ of the average background were removed to reduce noise contributions. Second, all intensity ratios measured at times when the Sun was not in the FOV of the Solar monitor were cut from the data set since the X-rays measured at this point are dominated by scattered radiation. The first cleaning criterion removed $\sim$50\% of the XRFS data, while the second cleaning criterion removed $\sim$25\% of the XRFS data. Furthermore, during orbit 16 (July 31, 1971), 67 and 73 (August 4, 1971) of Apollo 15, and orbit 19 (April 21, 1972) of Apollo 16, the Solar spectrum hardened, causing a decrease in all intensity ratios \citep{adler1972apollo, bielefeld1977imaging}. These orbits (4.7\% of total data) were taken out of the data set.  

Since the Solar spectrum was softer during the Apollo 16 mission (see Sect.~\ref{sec:dataprocessing}) with respect to that during the Apollo 15 mission, correction factors for the intensity ratios measured by Apollo 16 can be determined (see also \citealt{bielefeld1977lunar}). The intensity ratio measurements of Apollo 15 and 16 of the same area on the Lunar surface can be compared by cross-matching the measurements and determining a linear fit, yielding correction factors of 0.95, 0.83, and 0.84 for the Apollo 16 Al/Si, Mg/Si, and Mg/Al measurements, respectively. Here we used the intensity ratios remaining after the cuts and before the mapping procedure further outlined in Sect.~\ref{sec:intensity_ratio_maps}. 

Detector \#1 operated in the extended mode covering 1.4--6.0 keV for about 25\% of the observation time (see Tab.\ \ref{tab:observations}). This mode enables one to measure the fluorescence from elements with higher atomic number elements, such as calcium (Ca) and iron (Fe). The K$\alpha$ line energies of Ca and Fe are 3.69 and 6.40 keV, respectively (e.g., \citealt{SCHOONJANS2011776}). The Fe line can be detected in channel 8, which covers energies >6 keV, and since the Fe line is broadened in the detector it can also partially be detected in channel 7. In using the same method as described above, maps of the intensity ratios, such as Ca/Si, Fe/Si, and Fe/Ca, can be constructed. Due to the limited amount of data taken in the extended mode, however, only a few percent of the Lunar surface was covered. Also, the count rates in the channels corresponding to Ca and Fe line intensities were low. The Ca/Si, Fe/Si, and Fe/Ca maps are therefore of low quality and are not discussed further.

\section{Intensity ratio maps}
\label{sec:intensity_ratio_maps}

We constructed Al/Si, Mg/Si, and Mg/Al intensity ratio maps of the Apollo 15 \& 16 missions using the new collimator response and CSM orbit reconstruction. To visualize the X-ray intensity maps, a grid on the Lunar surface was constructed with a resolution of 0.1 degrees ($\sim$3 km) in longitude and latitude. The FOV was determined using the altitude at each 8 sec measurement and the newly determined collimator response. The rotation of the XRFS instrument with respect to the Lunar surface was also taken into account. For the creation of the Lunar maps, we used an equirectangular projection. We mapped the measured X-ray intensities onto the Lunar surface on each grid point and weighted the values with the transmission percentage. To include the 8 sec integration time, the transmission was smeared out over the distance traveled by the CSM, which was determined by interpolating the travel time between two following X-ray measurements. Finally, a weighted mean was determined for each point on the Lunar surface grid. Any points with only a single or no intensity ratio measurement with a higher than 20\% transmission percentage have been excluded from the map.

The Al/Si intensity ratios on the Lunar surface measured by Apollo 15 and 16 are shown in Fig.\ref{fig:alsi_map}. The map is projected onto an optical image of the Moon to investigate the location of certain geological features. The names of the known geological features are given in Appendix C in Fig.~\ref{fig:map_names}. It is apparent from the map that the Lunar mare (dark) regions typically have lower Al/Si ratios than the Lunar highland (light) regions.
The 1$\sigma$ uncertainties of the Al/Si intensity ratios are displayed in the bottom panel of Fig.~\ref{fig:alsi_map} and range from 0.01 to 0.36. It is important to note that the uncertainties on the Al/Si intensity ratios are especially high at the start and end of each track. This is likely to be caused by scattered light being more prominent than the fluorescent light due to Solar X-rays reaching the Lunar surface with a large angle and thus the total X-ray intensity being lower. Also, the data obtained near the terminator are less reliable due to poor counting statistics \citep{yin1993x}.

To quantify the distribution of the intensity ratios, we investigated the values as they were calculated and after they were stored on a grid of the Moon surface (i.e., "mapping"). The Al/Si distributions are shown in Fig.\ \ref{fig:double_peak_al_si}. The values before mapping are the intensity ratios calculated from each 8 sec measurement by Apollo 15 \& 16 remaining after the applied data cuts (see Sect. \ref{chemistry}). The intensity ratios after mapping are the mean intensity ratios on each grid point determined using the response of the collimator. Here it can be seen that after mapping, the Al/Si intensity ratio distribution is bimodal, which indicates the Lunar maria and highlands are clearly distinguishable, whereas before mapping this distinction could not be made.  

The Mg/Si and Mg/Al intensity ratio maps are shown in Fig.~\ref{fig:mg_si_mg_al_maps}. From these maps, it can be seen that the Lunar mare regions are characterized by higher Mg/Si and Mg/Al intensity ratios than the Lunar highlands. 
The intensity ratio distributions of Al/Si, Mg/Si, and Mg/Al with and without the correction factors for Apollo 16 (Section 2.3) are shown in Fig.\ \ref{fig:freq_distributions}. The corrections (right panel) result in additional peaks appearing in the Mg/Al and Mg/Si distributions, which is due to the relatively low values measured for Mg/Si and Mg/Al of the highland area between $\sim$100-150 degrees longitude by Apollo 16. The relative peak height in the bimodal distribution of Al/Si and Mg/Al is caused by the relative amount of maria and highlands covered by the CSM. A bimodal Gaussian distribution fit gives mean Al/Si values with 1$\sigma$ standard deviations of $0.54\pm0.07$ and $0.76\pm0.07$ with an Apollo 16 correction and $0.54\pm0.07$ and $0.78\pm0.08$ without corrections. The Mg/Al intensity ratio distributions are approximately bimodal with mean Mg/Al values of $0.54\pm0.17$ and $1.07\pm0.13$ with corrections and $0.57\pm0.17$ and $1.08\pm0.12$ without corrections. The Mg/Si intensity ratio is described by a single Gaussian distribution only, with a mean ratio of $0.47\pm0.13$ with corrections and $0.50\pm0.13$ without corrections. The Mg/Si data contain relatively more noise than the Al/Si data (see \citealt{Clarkphdthesis}), which may contribute to the less apparent distinction between the geological features in the Mg/Si map. Finally, it should be noted that channel 1 of Detector \#1 was noisy (see \citealt{Clarkphdthesis}), which may contribute to the uncertainty in the intensity ratios.

\begin{table*}
\caption{Lunar rock samples concentration ratios obtained during Lunar missions compared to average observed X-ray fluorescence intensity ratios in this work.}      
\label{tab:lunar_samples}      
\centering  
\resizebox{\textwidth}{!}{
\begin{tabular}{c c c c c c c c c c}  
\hline\hline       
Feature & Mission & Ref. & Rock type & \multicolumn{3}{c}{Concentration ratio} & \multicolumn{3}{c}{Observed intensity ratio*} \\
 & & & & Al/Si & Mg/Si & Mg/Al & Al/Si & Mg/Si & Mg/Al \\ 
\hline
Mare Tranquillitatis & Apollo 11 & [a] & High K rocks & 0.23 & 0.24 & 1.04 & 0.52 $\pm$ 0.10 & 0.62 $\pm$ 0.15 & 1.15 $\pm$ 0.24 \\ 
 &  & & Low K rocks & 0.29 & 0.23 & 0.79 & & & \\
 & Surveyor V & [b] & Regolith & 0.35 &  &  &  & \\
Mare Fecunditatis & Luna 16 & [c] & Rocks (average) & 0.35 & 0.21 & 0.6 & 0.58 $\pm$ 0.13 & 0.51 $\pm$ 0.14 & 0.89 $\pm $ 0.29\\ 
Hadley-Apennines & Apollo 15 & [d] & Fines (15021-80) & 0.34 & 0.28 & 0.85 & 0.53 $\pm$ 0.19 & 0.64 $\pm$ 0.20 & 1.26 $\pm$ 0.30 \\ 
Descartes Highlands & Apollo 16 & [d] & Fines (61161-7) & 0.67 & 0.18 & 0.26 & 0.81 $\pm$ 0.11 & 0.40 $\pm$ 0.14 & 0.49 $\pm$ 0.20 \\ 
Taurus-Littrow & Apollo 17 & [d] & Soil (72701-37) & 0.52 & 0.28 & 0.54 & 0.55 $\pm$ 0.09 & 0.55 $\pm$ 0.12 & 1.01 $\pm$ 0.26 \\ 
\hline
\hline 
\end{tabular}} 

References:
[a] \cite{levinson1970GeCAS...1.....L}, 
[b] \cite{turkevich1967Sci...158..635T},
[c] \cite{vinogradov1971preliminary}, 
[d] \cite{Wanke1973LPSC....4.1461W}. \\
* Obtained in this work 
\end{table*}

\section{Discussion \& conclusion}
\label{sec:discussion}

We performed a re-evaluation of the Lunar orbit observations done by the X-ray Fluorescence Spectrometers onboard Apollo 15 \& 16 using up-to-date knowledge of the detector, filters, and collimator transmission, as well as the CSM trajectory. This has resulted in updated Lunar surface maps of Al/Si, Mg/Si, and Mg/Al intensity ratios (see Fig.\ref{fig:alsi_map} and \ref{fig:mg_si_mg_al_maps}). These maps reveal Al-rich highlands and relatively Mg-rich maria, showing an inverse relationship between Al/Si and Mg/Si intensity ratios between the two types of terrain, as also stated in previous work (e.g., \citealt{adler1972apollo, apollo16preliminary, adlerapollo15preliminary1972}). We find, however, that our overall determined Al/Si intensity ratio values are $\sim$30\% lower than previously published (e.g., \citealt{adler1972apollo, bielefeld1977imaging, andre1977lunar}). For example, in averaging the values in the region of Mare Tranquillitatis, we find an Al/Si intensity ratio of 0.52, whereas \cite{bielefeld1977imaging} found ~0.75 and \cite{adler1972apollo} found 0.81$\pm$0.13. 
Our intensity ratio maps are in agreement with low-resolution global Lunar surface maps derived from the Chang'E-2 X-ray spectrometer \citep{Dong2016RAA....16....4D}. 

The differentiation between mare and highland regions is clearly seen in the bimodal intensity ratio distribution (see Fig. \ref{fig:freq_distributions}). Furthermore, Lunar mare regions in general are characterized by higher Mg/Si and Mg/Al intensity ratios than the Lunar highlands, although this is less clear for the Mg/Si intensity ratio maps
(see also, e.g., \citealt{clark1978utilization, clarkhawke1981}). The comparatively higher Al/Si intensity ratio values in the highlands are due to the presence of anorthosite type rocks (mostly plagioclase feldspar). The mare regions, however, are characterized by high Mg/Si intensity ratio values compared to the highland regions due to relatively Mg-rich basaltic lavas, typically abundant in the mineral pyroxene (see, e.g., \citealt{Heiken1991S&T....82..498H}).

There are multiple methods that can be used to convert the observed intensity ratios into concentration ratios. Firstly, previous work has shown that there is a direct relationship between these measured intensity ratios and the concentration of the bulk soil \citep{Hubbard1980luhc.conf..457H}. Therefore, concentration ratios can be derived by combining remote sensing observations with Lunar sample investigations (see, e.g., \citealt{adler1972apollo, trombka1977NASSP.370..153T}), which can be directly determined through compositional analyses of samples from Lunar landing sites (e.g., \citealt{Heiken1991S&T....82..498H}). Secondly, the Solar spectrum can be reconstructed by comparing Solar observations in multiple energy bands to theoretical solar emission models, which can be used to determine the expected fluorescence. The relation between the hardness of the solar spectrum and expected fluorescence can then be used to correct the measured intensity ratios to concentrations (see e.g., \citealt{hubbard1978Solar, clark1978utilization, clark1997remote}). During the Apollo 15\&16 missions, Solar observations were obtained by SOLRAD 10, which can also be used for this analysis (see e.g., \citealt{clark1978utilization}). Finally, in the lab on the ground, element line intensity ratios can be directly related to element concentration ratios by experimentally measuring standards. We have not been able to find such calibration reports for the XRFS.

To obtain an estimate of the concentration ratios derived from our work, we compared our observed intensity ratios to concentration ratios obtained from Lunar samples from various missions in Tab.\ \ref{tab:lunar_samples}. The comparison between intensity ratio and concentration values can be used to generate an approximate calibration for these data. These values have been plotted in Fig.~\ref{fig:rock_conversion} and are used to derive a linear relation between the measured intensity ratio and concentration ratio. Scaling the intensity ratios with these linear fits yields rough estimates for the concentration ratio of the mare and highland regions of 0.34$\pm$0.11 and 0.71$\pm$0.13 for Al/Si and 0.18$\pm$0.19 and 0.77$\pm$0.15 for Mg/Al, respectively. Again, no distinction can be made between the mare and highlands for the Mg/Si distribution; therefore, the distribution is described by a single Gaussian with a mean of 0.19$\pm$0.08. 

In future studies an analysis of the measured Solar radiation itself can be used to construct a more accurate comparison between the Apollo 15 and Apollo 16 data, as well as inter-orbital variations during the individual missions {(see, e.g., \citealt{clark1978utilization, bielefeld1977imaging})}. These inter-orbital variations are largest for the Mg/Si ratio and are also partly caused by variations in surface roughness \citep{hubbard1977reevaluation}. In addition, the differences in the instrument response and energy gain of Apollo 15 and 16  need further investigation to determine the correction needed for the Apollo 16 data. Also, the contribution from scattered X-ray photons needs to be disentangled from the pure fluorescent X-rays; this cannot be obtained from the Lunar data alone. The effect of scattered radiation is greatest on the Mg/Si intensity ratio since the exciting Solar flux strongly increases with energy and the Mg and Si line energies deviate the most \citep{bielefeld1977lunar}. These extra corrections, together with a calibration to convert intensities to a concentration, will result in improved concentration maps, which can then finally be used to study the concentration ratios over an extended area of the Lunar surface and contribute to our current understanding of the Lunar surface geology and formation. 

\begin{acknowledgements}

{The authors thank the referee for the useful suggestions and comments. We are grateful to Bernard Foing, Juan García Redondo, and Francesca McDonald for the useful suggestions and discussions regarding this work. AJG and YM thank ESA/ESTEC and ESA/ESAC, respectively, for its hospitality during an internship. AE and MK acknowledge support from the ESTEC Faculty Visiting Scientist Programme. YM also acknowledges support from ESA through the Faculty of the European Space Astronomy Centre (ESAC) - Funding reference ESAC-357. EK thanks David R.\ Williams for making available the restored binary offline tape datasets from the NSSDC Archive and Gerald K.\ Austin for providing the XRFS Prototype Unit.}

\end{acknowledgements}

\bibliographystyle{aa}
\bibliography{bibliography}

\begin{appendix}

\section{Filter thickness}
\label{sec:appendix_filterthickness}

The thickness of the Mg and Al filters, as well as that of the Be windows, is not entirely clear from the literature as different values have been published (e.g., \citealt{adler1972apollo, adler1975apollo, Clarkphdthesis, yin1993x}). The Mg and Al filter thickness is, for example, stated to be in the range of 5.08--12.7 $\mu$m in \cite{adler1972apollo}, while values of 5.1 and 6.4 $\mu$m are given by \cite{Clarkphdthesis} and \cite{yin1993x}, respectively. The Be window thickness is quoted as $\sim$25 $\mu$m in most work (e.g., \citealt{adler1972apollo, adler1975apollo, Clarkphdthesis, yin1993x, clark1997remote}); however, for example, it is 40 $\mu$m in \cite{adler1973lunar}. It should be noted that the resulting intensity ratios highly depend on the assumed window and filter thickness since this determines the amount of X-ray transmission.

\begin{figure}[h]
        \includegraphics[width=\columnwidth]{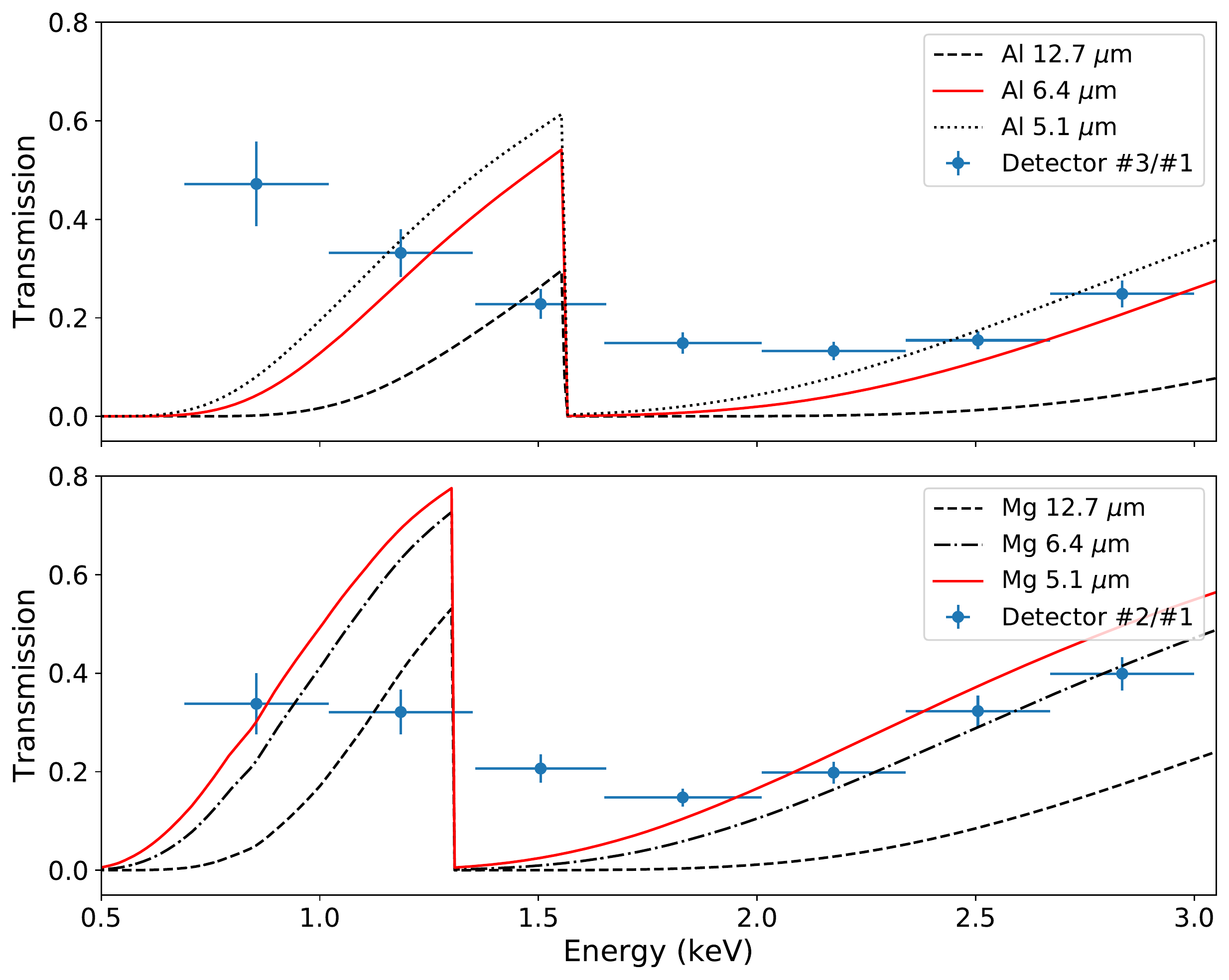}
    \caption{Transmission of the Al (Detector \#3/\#1, top) and Mg (Detector \#2/\#1, bottom) filter when observing Scorpius X-1 during the trans-Earth coast flight of Apollo 15. Filter transmission of Al (top) and Mg (bottom) is given as a function of energy for different thicknesses obtained from \url{http://henke.lbl.gov/optical_constants/}  \citep{HENKE1993181}.}
    \label{fig:trans_sco}
\end{figure}

An attempt to determine the correct thickness of the Al and Mg filter can be made. The filter transmission can be estimated from the XRFS observations of, for example, the bright X-ray source Scorpius X-1 during the trans-Earth coast flight obtained by Apollo 15 on August, 5 1971. By using the ratios of overall detected counts in each channel of Detectors \#3 and \#1 (Al/no filter) and Detectors \#2 and \#1 (Mg/no filter), the effect of the Be window was divided out  when the thickness of the Be window was assumed to be the same for all three detectors. The background in each channel during the trans-Earth coast flight was determined from observations taken before and after the Scorpius X-1 measurement when the cover was open. The resulting transmission as a function of channel energy is shown in Fig.\ \ref{fig:trans_sco}. This method does not favor a thickness of 5.1 or 6.4 $\mu$m for either filter, but it does seem to rule out a thickness of 12.7 $\mu$m. We note that the Al transmission in the first channel is too high for either filter thickness, which may be due to instrumental noise in that channel of Detector \#1 (see \citealt{Clarkphdthesis}).
Furthermore, we measured the filter thickness on the XRFS prototype unit (see Sect.~\ref{sec:dataprocessing}); however, due to exposure to air during their storage, these filters have been affected, hence yielding inaccurate measurements. 
In this work, we therefore adopted the Be window thickness and Al \&\ Mg filters thickness (see Tab.\ref{tab:var_trans}) as published in \cite{clark1978utilization}, \cite{Clarkphdthesis} and \cite{yin1993x}.\\

\section{Filter transmission}
\label{sec:appendix_filtercalculation}

In this section, the determination of Al, Si, and Mg intensities as discussed in Sect.\ \ref{sec:dataprocessing} is further explained. Equation \ref{eq:transmission} in its complete form is given by {(see, e.g., \citealt{adler1972apollo})}:

\begin{equation}
\begin{aligned}
\label{eq:flux_equation}
    G_1 &= I^T_{\text{Mg}} T_{\text{Mg}}^{Be}F_{1\text{Mg}} + I^T_{\text{Al}}T^{Be}_{\text{Al}}F_{1\text{Al}} + I^T_{\text{Si}}T^{Be}_{\text{Si}}F_{1\text{Si}}, \\
    G_2 &= I^T_{\text{Mg}} T_{\text{Mg}}^{Be}F_{2\text{Mg}} + I^T_{\text{Al}}T^{Be}_{\text{Al}}F_{2\text{Al}} + I^T_{\text{Si}}T^{Be}_{\text{Si}}F_{2\text{Si}}, \\
    G_3 &= I^T_{\text{Mg}} T_{\text{Mg}}^{Be}F_{3\text{Mg}} + I^T_{\text{Al}}T^{Be}_{\text{Al}}F_{3\text{Al}} + I^T_{\text{Si}}T^{Be}_{\text{Si}}F_{3\text{Si}}.
\end{aligned}
\end{equation}

In these equations, G$_{i}$ are the background subtracted counts in detector $i$ for $i=1,2,3$. Furthermore, I$^T$ is the total flux at the Mg, Al, or Si K$\alpha$ line energy, T$^{Be}_{j}$ is the transmission efficiency of the Be window, and F$_{ij}$ is the transmission efficiency of each detector's $i$ at the Al, Si, and Mg K$\alpha$ line energies (given by $j$). The transmission efficiencies of the Be, Mg, and Al filter are given by the exponential function $e^{-\mu_i \rho_i d_i}$, with $\mu$ being the absorption cross section, $\rho$ being the density, and $d$ being the thickness of the filter (see Equations 2, 3, and 4). Filling out the values of these constants as given in Tab.\ \ref{tab:var_trans} results in a filter matrix M of

\begin{equation}\label{filter_matrix_eq}
\begin{aligned}
M &= 
\begin{bmatrix}
T_{\text{Mg}}^{Be}F_{1\text{Mg}} & T^{Be}_{\text{Al}}F_{1\text{Al}} & T^{Be}_{\text{Si}}F_{1\text{Si}} \\ T_{\text{Mg}}^{Be}F_{2\text{Mg}} & T^{Be}_{\text{Al}}F_{2\text{Al}} & T^{Be}_{\text{Si}}F_{2\text{Si}} \\ T_{\text{Mg}}^{Be}F_{3\text{Mg}} & T^{Be}_{\text{Al}}F_{3\text{Al}} & T^{Be}_{\text{Si}}F_{3\text{Si}}
\end{bmatrix} \\ &= 
\left[\begin{matrix}

e^{-(\mu_{Be}^{Mg} \rho_{Be} d_{Be})} & e^{-(\mu_{Be}^{Al}\rho_{Be} d_{Be})}  \\ 
e^{-(\mu_{Be}^{Mg} \rho_{Be} d_{Be}+\mu_{Mg}^{Mg} \rho_{Mg} d_{Mg})} & e^{-(\mu_{Be}^{Al} \rho_{Be} d_{Be}+\mu_{Mg}^{Al} \rho_{Mg} d_{Mg})}  \\ 
e^{-(\mu_{Be}^{Mg} \rho_{Be} d_{Be}+\mu_{Al}^{Mg} \rho_{Al} d_{Al})} & e^{-(\mu_{Be}^{Al} \rho_{Be} d_{Be}+\mu_{Al}^{Al} \rho_{Al} d_{Al})} 
\end{matrix}\right.\\
&\qquad\qquad
\left.\begin{matrix}
    e^{-(\mu_{Be}^{Si} \rho_{Be} d_{Be})} \\
    e^{-(\mu_{Be}^{Si} \rho_{Be} d_{Be}+\mu_{Mg}^{Si} \rho_{Mg} d_{Mg})} \\
    e^{-(\mu_{Be}^{Si} \rho_{Be} d_{Be}+\mu_{Al}^{Si} \rho_{Al} d_{Al})}
\end{matrix}\right] \\ &=
\begin{bmatrix}
0.2373 & 0.4222 & 0.5857 \\ 
0.1520 & 0.0113 & 0.0496 \\ 
0.0774 & 0.2073 & 0.0027
\end{bmatrix}
.
\end{aligned}
\end{equation}

It is important to note that this filter matrix is almost identical, as can be deduced from values given by \cite{yin1993x}.
Finally, the equation given by

\begin{equation}
\begin{aligned}
    \begin{bmatrix}
    G_1 \\
    G_2 \\
    G_3
    \end{bmatrix} &=
    M \begin{bmatrix}
    I^T_{Mg} \\
    I^T_{Al} \\
    I^T_{Si}
    \end{bmatrix}
\end{aligned} \\
\end{equation}

was solved to obtain $I^T_{Mg}$, $I^T_{Al}$, and $I^T_{Si}$. Once these three intensities were obtained, the Al/Si, Mg/Si, and Mg/Al intensity ratios were given by $I^T_{\text{Al}}/I^T_{\text{Si}}$, $I^T_{\text{Mg}}/I^T_{\text{Si}}$, and $I^T_{\text{Mg}}/I^T_{\text{Al}}$, respectively. The residuals obtained from the least square inversion method are negligible. Instead, Poisson statistics (square root of the number of counts) were used to estimate the uncertainty in the intensity ratios.

\section{Rock concentration and geological features}

\begin{figure*}[h]
\centering
   \includegraphics[width=\textwidth, trim={0.0cm 0cm 0cm 0.0cm}, clip]{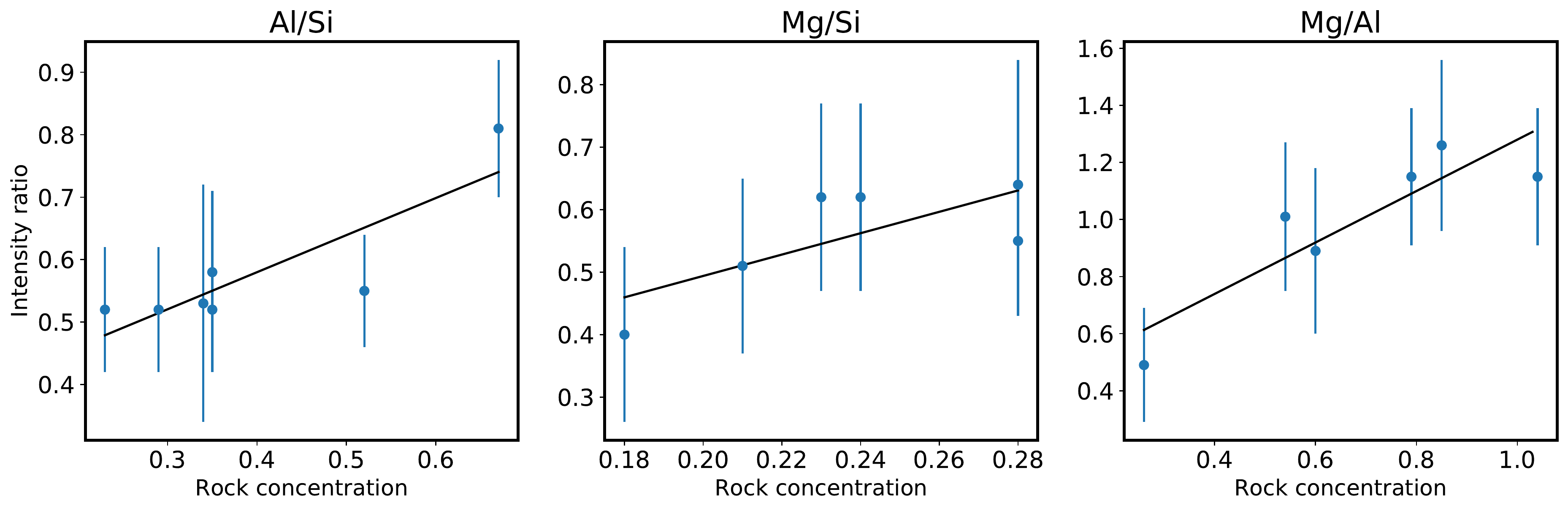}
     \caption{Intensity ratio as a function of the measured Lunar rock concentration of Al/Si (left), Mg/Si (middle), and Mg/Al (right) from Tab.~\ref{tab:lunar_samples} with the black line showing the optimal linear fits.}
     \label{fig:rock_conversion}
\end{figure*}

\begin{figure*}
\centering
   \includegraphics[width=0.95\textwidth, trim={0.0cm 2.5cm 0cm 2.0cm}, clip]{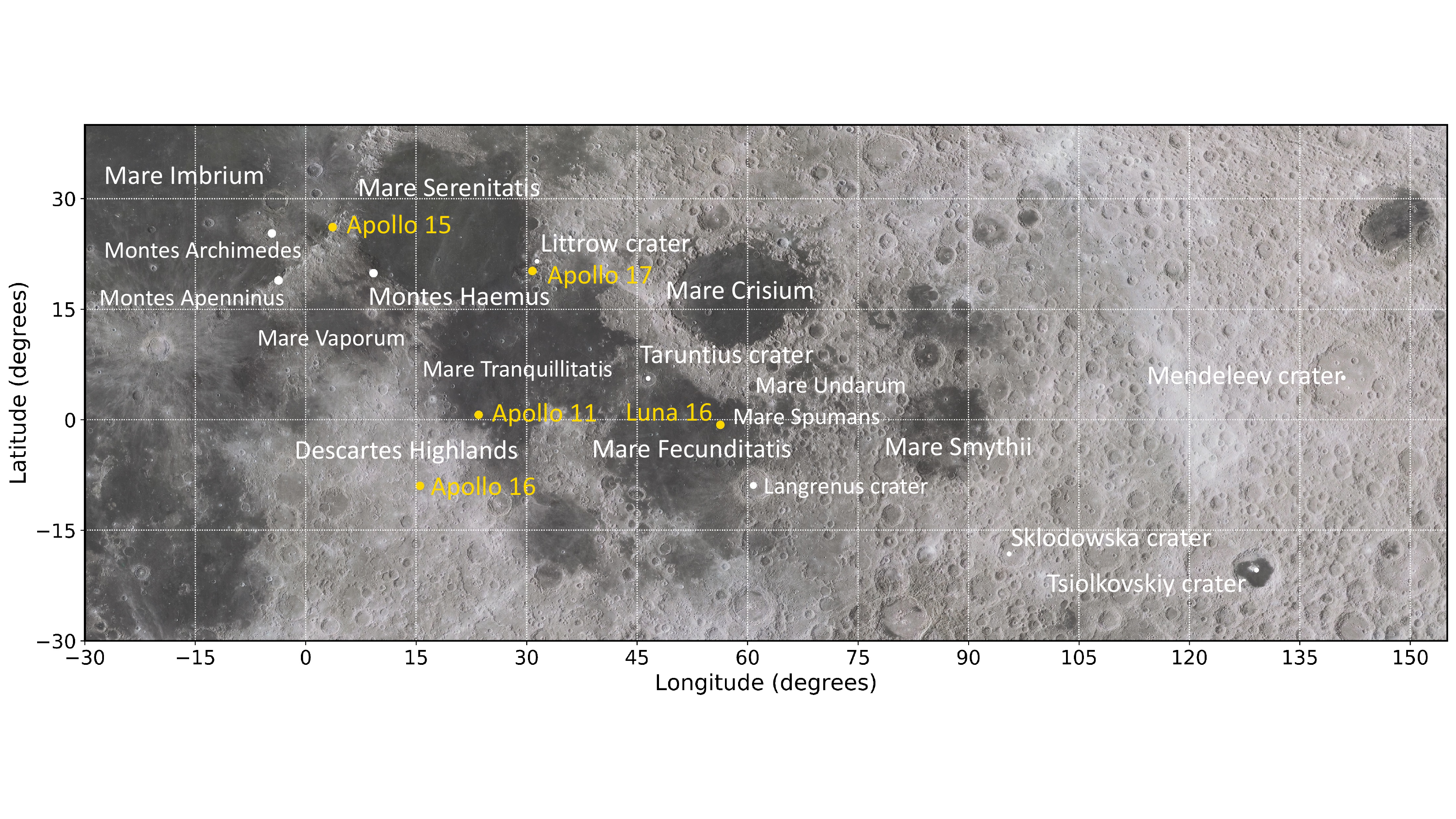}
     \caption{Map of Lunar surface with names of geological features. The landing site of Apollo 11, 15, 16, 17, and Luna 16 are given in yellow.}
     \label{fig:map_names}
\end{figure*}

Fig.~\ref{fig:rock_conversion} shows the intensity ratios versus rock concentration ratios from Tab.~\ref{tab:lunar_samples}. These linear fits were used to give an estimate for the mean measured concentration ratios (see Sect.~\ref{sec:discussion}). Fig.\ \ref{fig:map_names} gives the geological names of the most prominent Lunar features covered by the Apollo 15 and 16 XRFS experiment. 

\end{appendix}

\end{document}